\documentclass[aps,prb,twocolumn,superscriptaddress,showpacs,preprintnumbers]{revtex4}

\usepackage{graphicx}

\begin{document}


\title{Intrinsic electronic superconducting phases \\ at 60 K and 90 K
in double-layer YBa$_2$Cu$_3$O$_{6+\delta}$}

\author{T. Honma}
\affiliation{Department of Physics, Asahikawa Medical College, Asahikawa, Hokkaido 078-8510, Japan}
\author{P.H. Hor}
\affiliation{Department of Physics and Texas Center for Superconductivity and Advanced Materials, University of Houston, Houston, TX77204-5005, U.S.A.}

\date{\today}

\begin{abstract}
We study superconducting transition temperature ($T_c$) of oxygen-doped double-layer high-temperature superconductors YBa$_2$Cu$_3$O$_{6+\delta}$ (0 $\le$ $\delta$ $\le$ 1) as a function of the oxygen dopant concentration ($\delta$) and planar hole-doping concentration ($P_{pl}$). We find that $T_c$, while clearly influenced by the development of the chain ordering as seen in the $T_c$ $vs.$ $\delta$ plot, lies on a universal curve originating at the critical hole concentration ($P_c$) = 1/16 in the $T_c$ $vs.$ $P_{pl}$ plot. Our analysis suggests that the universal behavior of $T_c$($P_{pl}$) can be understood in terms of the competition and collaboration of chemical-phases and electronic-phases that exist in the system.  We conclude that the global superconductivity behavior of YBa$_2$Cu$_3$O$_{6+\delta}$ as a function of doping is electronically driven and dictated by pristine electronic phases at magic doping numbers that follow the hierarchical order based on $P_c$, such as 2 $\times$ $P_c$, 3 $\times$ $P_c$ and 4 $\times$ $P_c$. We find that there are at least two intrinsic electronic superconducting phases of $T_c$ = 60 K at 2 $\times$ $P_c$ = 1/8 and $T_c$ = 90 K at 3 $\times$ $P_c$ = 3/16. 
\end{abstract}

\pacs{74.25.Fy, 74.25.Dw, 74.62.Dh, 74.72.Bk}


\maketitle

It has become clear in recent years that various physical properties of high temperature superconductors (HTS) are manifestations of a complex electronic texture of intrinsic electronic inhomogeneities due to dopants and, more importantly, electronic phase separation (EPS). In this report we show that the key to understand the underlying electronic texture and the corresponding superconducting properties are electronic phases that exist at magic planar doping concentrations (hole content per CuO$_2$ plane, $P_{pl}$) $P_{pl}$ = $m$/$n^2$, where both $m$ and $n$ are positive integers with 4$m$ $\leq$ $n^2$ and $P_{pl}$ is determined based on a universal hole scale. Studies of the electronic phase diagram under ambient and high-pressure in cation (Sr) and anion (O) co-doped polycrystalline La$_{2-x}$Sr$_x$CuO$_{4+\delta}$ (CD-La214) revealed that there are two intrinsic electronic superconducting phases with superconducting transition temperature $T_{c1}$ = 15 K and $T_{c2}$ = 30 K. \cite{lor02} The far-infrared charge dynamics studies on the $T_{c1}$ and $T_{c2}$ phases indicated that they are very peculiar electronic phases which have very small amount ($<$ 1 $\%$ of total doped hole) of free holes moving in otherwise pinned two-dimensional (2D) electronic lattice. The $T_{c1}$ and $T_{c2}$ phases start at the critical \textquotedblleft magic\textquotedblright\ planar hole-doping levels $P_{pl}$ = 1/16 $\equiv$ $P_c$ and $P_{pl}$ = 1/8 = 2 $\times$ $P_c$, respectively. \cite{kim01} The existence and the clear competition of $T_{c1}$ and $T_{c2}$ phases observed in the pure cation (Sr)-doped polycrystalline La$_{2-x}$Sr$_x$CuO$_4$ (SrD-La214) indicated that, independent of the nature of the dopants, these intrinsic $T_c$'s phases are energetically favored electronic phases that exist in the CuO$_2$ planes. \cite{hor02} Most recently magnetic studies of SrD-La214 single crystals confirmed the existence of the $T_{c1}$ and $T_{c2}$ transitions and the onset of the superconducting transition temperature were surprisingly robust with little field dependence up to 5 T. \cite{don06} These \textquotedblleft 2D square electronic lattices\textquotedblright\ formed at magic doping concentrations are the most fundamental building blocks of electronic states, the pristine electronic phase (hereafter PEP), for the understanding of both normal and superconducting properties of HTS. While three-dimensional ordered PEP's seem to be firmly established in single-layer La214 system, \cite{don06} it is not clear how will PEP manifest themselves in other HTS. In this report we show that, indeed, in pure anion (O) doped double-layer YBa$_2$Cu$_3$O$_{6+\delta}$ (OD-Y123) system there exist at least two intrinsic $T_c$'s of $T_c$ = 60 K and $T_c$ = 90 K that are based on PEP's at $P_{pl}$  = 2/16 = 2$P_c$ and 3/16 = 3$P_c$, respectively. Furthermore we find that the famous \textquotedblleft 60 K-plateau\textquotedblright\ in the electronic phase diagram of OD-Y123 can be naturally explained by the EPS of PEP's

In order to sort out the PEP's in OD-Y123 system we need to take care of the complications due to sensitive dependences of $T_c$ on both the amount and arrangement of oxygen dopants. \cite{jor90,vea90} The oxygen dopants tend to form long Cu-O chain ordering along $b$-axis that results in energetically favored meta-stable superstructures consisting of alternative arrangements of full Cu-O chain (full-chain) and O-vacancy chain (empty-chain). \cite{and99} It was proposed that there are, starting from complete full-chain-ordering ortho-I phase, ortho-II (-II), -III (-III$^*$) and -IV (-IV$^*$) chemical phases will have single, double and triple full-chains (empty-chains) between any two empty-chains (full-chains), respectively. \cite{and99} It is well known that in the $T_c$ $vs.$ $P_{pl}$ phase diagram of OD-Y123 there are two prominent plateaus located at $T_c$ $\sim$60 K (60K-plateau) and $\sim$90 K (90K-plateau). It was generally assumed that 60 K and 90 K phases corresponded to the ortho-II and ortho-I chemical phases, respectively. \cite{pou91,cal97} There were two possible origins of 60K-plateau proposed: one is the chemical phase separation \cite{bey89} and the other is purely electronic in origin. \cite{seg01} Recent theoretical study indicated that chemical phase arguments based on chain ordering alone can not account for the observed 60K-plateau. In stead, a chain-ordering induced charge imbalance model was used to account for the 60K-plateau. \cite{zal06} The electronic scenario attributed 60K-plateau to the well-known 1/8 anomaly identified in the SrD-La214. \cite{ako98} However the analysis was, unfortunately, based on a questionable planar hole scale. \cite{obe92,kni99,hon04} The problem concerning the origin of the $T_c$ = 60 K and 90 K phases and the associated plateaus in OD-Y123 are still unresolved. In this report we show that the electronic phase diagram of OD-Y123 can be understood in terms of the competition and the collaboration among the chemical phases and the PEP's.

\begin{figure}[b]
\includegraphics[scale=1.1]{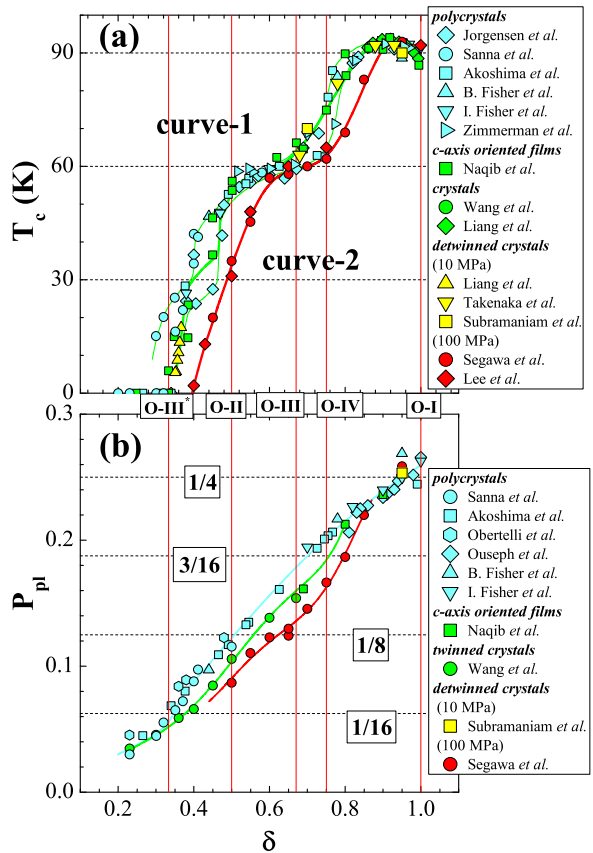}
\caption{\label{fig1} (Color online) (a) Superconducting transition temperature ($T_c$) as a function of excess oxygen content ($\delta$) for YBa$_2$Cu$_3$O$_{6+\delta}$. \cite{jor90,ako98,san03,san05,bfi93,ifi94,naq05,wan01,lia98,seg01,lia02,lee05,sub95,tak94} The solid curves are guide to the eyes. The solid and dotted vertical lines show the idealized chemical phases of the ortho-I (O-I), -II (O-II), -III (O-III), -IV (O-IV) and ortho-III$^*$ (O-III$^*$). (b) Hole concentration ($P_{pl}$) as a function of $\delta$ for YBa$_2$Cu$_3$O$_{6+\delta}$. \cite{ako98,obe92,san05,bfi93,ifi94,ous90,naq05,wan01,seg01,lee05} The dotted horizontal lines show the magic number of the hierarchy of $P_c$ = 1/16. The vertical lines show the idealized chemical phases of the ortho-I (O-I), -II (O-II), -III (O-III), -IV (O-IV) and ortho-III$^*$ (O-III$^*$). $P_{pl}$ of all data, except of Akosima $et$ $al.$'s work, are directly determined from the reported $S^{290}$ by using of the universal scale in ref.\ \onlinecite{hon04}. Akoshima $et$ $al$. reported the $\delta$ $vs.$ the hole concentration determined from their own $S^{290}$ by using the scale of ref. \onlinecite{tal95}. We plotted the re-determined $P_{pl}$ by converting their used scale into our scale. The solid curves are guide to the eyes.}
\end{figure}

A universal hole scale based on the thermoelectric power at 290 K ($S^{290}$) was constructed and used for comparing various physical properties in HTS. \cite{hon04} It was shown that both normal and superconducting properties can be compared consistently with systematic doping dependences among different HTS. \cite{hon04,hon06} We analyze the reported $T_c$($\delta$) data of OD-Y123 as a function of $P_{pl}$ determined by the universal scale. We find that $T_c$($P_{pl}$) lies on one universal curve originating at the critical hole concentration $P_c$ = 1/16, while $T_c$($\delta$) lies on several curves which strongly depend on the development of the chain ordering. Further, we also find that the robust 60K- and 90K-plateaus appear at $P_{pl}$ $\sim$1/8 = 2$P_c$ and $\sim$ 3/16 = 3$P_c$, respectively and the superconductivity is always suppressed beyond $P_{pl}$ = 1/4 = 4$P_c$. We extracted the $T_c$ data from the published paper, irrespective of the definition of $T_c$. \cite{jor90,ako98,san03,san05,bfi93,ifi94,naq05,wan01,lia98,seg01,lia02,san04,pop97,lee05} A sample's $P_{pl}$ was directly determined from the reported $S^{290}$ data by using the universal scale. \cite{ako98,obe92,san05,bfi93,ifi94,naq05,wan01,seg01,ous90,pop97,lee05} For samples with either $\delta$ or $T_c$ reported but not $S^{290}$, the $P_{pl}$ of the sample were determined from the $P_{pl}$ $vs.$ $\delta$  or the $P_{pl}$ $vs.$ $T_c$ relation as discussed below. \cite{san03,lia98,lia02,san04,uem89,pum90,zim95}

\begin{figure}[b]
\includegraphics[scale=1.1]{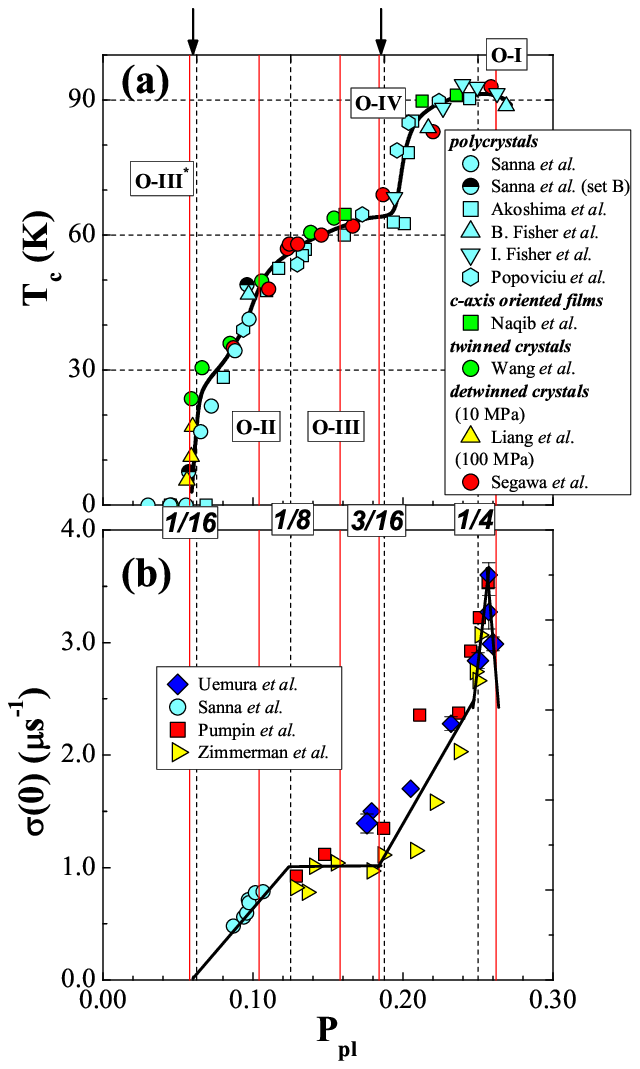}
\caption{\label{fig2} (Color online) (a) $T_c$ $vs.$ $P_{pl}$ for YBa$_2$Cu$_3$O$_{6+\delta}$ (OD-Y123). \cite{ako98,san05,bfi93,ifi94,naq05,wan01,lia98,seg01,san04,pop97,lee05} The $P_{pl}$ for $T_c$ data with $\delta$ value alone by Liang $et$ $al$. \cite{lia98} was determined from the $P_{pl}$ $vs.$ $\delta$ plot of Fig. 1(b). The second data set by Sanna $et$ $al$. (set B) was reported in ref.\ \onlinecite{san04}. According to ref.\ \onlinecite{san04}, we can extract the lower and upper bound of $\delta$ and $T_c$ for their OD-Y123 samples. The ($\delta$, $T_c$) are (0.32, 7 K) and (0.42, 49 K). We plotted these two points by determining $P_{pl}$ from the $\delta$ value. All other $P_{pl}$ were directly determined from the reported $S^{290}$. (b) $N_S$ $vs.$ $P_{pl}$ for the OD-Y123. \cite{san04,uem89,pum90,zim95} $P_{pl}$ for $\sigma$(0) data with $T_c$ value alone by Sanna $et$ $al$. \cite{san04} were determined from the $P_{pl}$ $vs$. $T_c$ plot of Fig. 2(a). All other $P_{pl}$ for $\sigma$(0) with $\delta$ and $T_c$ \cite{uem89,pum90,zim95} were determined from the $P_{pl}$ $vs.$ $\delta$ plot of Fig. 1(b). In the Fig. 2(a) and 2(b), the vertical broken lines show the magic number of the hierarchy of $P_c$ = 1/16. The vertical lines show the idealized chemical phases of the ortho-I (O-I), -II (O-II), -III (O-III), -IV (O-IV) and ortho-III$^*$ (O-III$^*$). The solid curves guide to the eyes.}
\end{figure}

In figure\ \ref{fig1}(a) we plot $T_c$ $vs.$ $\delta$ for OD-Y123. The $T_c$($\delta$) behavior can be roughly separated into two curves with small loop structures for $\delta$ $<$ 0.47 and 0.7 $<$ $\delta$ $<$ 0.8. Curve-1 is $T_c$($\delta$) behavior corresponding to the polycrystals, \cite{jor90,ako98,san03,bfi93,ifi94} crystals \cite{wan01,lia98,lia02} and $c$-axis oriented films. \cite{naq05} Curve-2 is $T_c$($\delta$) behavior as observed in the crystals detwinned by a uniaxial pressure of 100 MPa \cite{seg01,lee05} applied along $a$($b$)-axis. For all the curve-1 samples, the $T_c$ appears at $\delta$ $\sim$ 0.35, reaches the 60K-plateau for 0.47 $<$ $\delta$ $<$ 0.65 and reaches the 90K-plateau for 0.82 $<$ $\delta$ $<$ 0.92. Finally, the $T_c$ slightly decreases for $\delta$ $>$ 0.92. For the curve-2 samples, the $T_c$ appears at $\delta$ $\sim$ 0.4, exhibits the 60K-plateau at 0.6 $<$ $\delta$ $<$ 0.75 and reaches the 90K-plateau at $\delta$ $\sim$ 0.9. The $T_c$($\delta$)-curve of the detwinned crystals is lower than that of the curve-1 samples. However, the $T_c$($\delta$)-curve of crystals detwinned by the uniaxial pressure of 10 MPa \cite{lia02,sub95,tak94} follows that of the curve-1 samples. This indicates that the artificially prepared very long chain ordering actually suppresses the superconductivity. All of above are intrisically consistent with the facts that the PEP's are 2D square lattice. Furthermore the report that the highest $T_c$ in La$_{2-x}$M$_x$CuO$_4$ (M = Nd, Ca, Sr) system is always observed in the tetragonal phase with flat and square CuO$_2$ planes is also consistent with the above picture. \cite{dab96} Clearly superconductivity of OD-Y123 is greatly affected by the level of the chain ordering and no systematic universal behavior can be inferred from the $T_c$ $vs.$ $\delta$ phase diagram depicted in figure\ \ref{fig1}(a).

We plot $P_{pl}$ $vs.$ $\delta$ in figure\ \ref{fig1}(b). Starting from $\delta$ $\sim$ 0.35 where $P_{pl}$ $\sim$ 1/16 = $P_c$, the $P_{pl}$($\delta$) curve separated into three curves that merge into a common curve for $\delta$ $>$ $\sim$0.85, where $P_{pl}$ $\sim$ 2/9. The upper, middle and lower curves are determined from the $S^{290}$ of polycrystals, \cite{ako98,obe92,san05,bfi93,ifi94,ous90} the in-plane $S^{290}$ of twinned crystals including $c$-axis oriented films, \cite{naq05,wan01} and the $S^{290}$ measured along $a$-axis ($S_a^{290}$) of crystals detwinned by the uniaxial pressure of 100 MPa, \cite{seg01} respectively. $P_{pl}$ increases monotonically with oxygen doping. The continuous increase in $P_{pl}$ is not consistent with the chemical phase separation. \cite{vea91} The common $P_{pl}$ curve for $\delta$ $>$ 0.85 suggests, in contrast to the charge imbalance model,\cite{zal06} that chain ordering has no influences on hole concentration in the CuO$_2$ planes. Accordingly, the $S^{290}$ scale in ref.\ \onlinecite{hon04} can also be used for $\delta$ $>$ 0.85 in OD-Y123. In the $\delta$-range from $\sim$0.35 to $\sim$0.85, the planar hole concentration at the same $\delta$ values is successively suppressed in the order of the polycrystals, twinned crystals and detwinned crystals with increasing length of chain ordering. Therefore, in contrast to the common belief, the perfect chemical phase with long chains does not favor the electronic state of HTS. Hereafter, if we only know the value of $\delta$ in the OD-Y123, we estimate the $P_{pl}$ for the polycrystals, twinned crystals and artificially detwinned crystals from the corresponding curve of $P_{pl}$ $vs.$ $\delta$ plot in Fig.\ \ref{fig1}(b)

In figure\ \ref{fig2}(a) we plot $T_c$ $vs.$ $P_{pl}$ for OD-Y123. \cite{ako98,san05,bfi93,ifi94,naq05,wan01,lia98,seg01,san04,pop97,lee05} Surprisingly, $T_c$ data of all samples collapsed into a single universal curve in the $T_c$ $vs.$ $P_{pl}$ plot. The superconductivity appeared at $P_{pl}$ $\sim$ 0.06 ($\sim$ $P_c$) that exhibits an extremely sharp $T_c$-jump to $\sim$30 K followed by a broad $T_c$ increase to 60 K at $P_{pl}$ $\sim$ 0.1. The 60K-plateau is observed between 0.12 ($\sim$ 2$P_c$) $<$ $P_{pl}$ $<$  0.19 ($\sim$ 3$P_c$). Further, the $T_c$ suddenly jumps to $\sim$90 K over $\sim$0.19 ($\sim$ 3$P_c$) and goes into the 90K-plateau for 0.21 $<$ $P_{pl}$ $<$ 0.25 ($\sim$ 4$P_c$). Finally, the $T_c$ decreases for $P_{pl}$ $>$ 0.25 ($\sim$ 4$P_c$). The phase diagram is characterized by fast $T_c$-jumps and much flatter $T_c$ regions in between. Noted that the sharp $T_c$-jumps occurred whenever the $P_{pl}$ of a chemical phase is almost identical to that of a PEP (See two arrows in Fig.\ \ref{fig2}(a)). In contrast, the broad $T_c$-increase at $\sim$1/8 seems to come from the mismatch between the $P_{pl}$ of chemical phases to 2$P_c$. The above observations suggest that the global behavior of the electronic phase diagram is dictated by the EPS of PEP's under the influences of chemical phases: the jump in $T_c$ is due to the matching of a PEP and a chemical phase and the flat region is a two PEP's region. There are some fine structures in the $T_c$ $vs.$ $P_{pl}$ plot; a \textquotedblleft hint\textquotedblright\ of a 30 K intrinsic phase and a 30K-plateau around $P_{pl}$ = $P_c$ and a small $T_c$-jump to $T_c$ $\sim$ 45 K occurred when ortho-II matched with magic doping concentration at $P_{pl}$ $\sim$ 1/9. It is also interesting to note that the intrinsic $T_c$'s increase as integer multiples of 30 K in the double-layer OD-Y123 in contrast to that of 15 K in the single-layer La214 system.

The above EPS picture for the 60K-plateau is further collaborated by the observation of a clear plateau of another intrinsic property, the superfluid density ($N_S$), of a superconductor in the same two phase region. We used $N_s$ estimated from the low temperature muon-spin relaxation ($\mu$SR) rate $\sigma$(0). In Fig.\ \ref{fig2}(b), we plotted $\sigma$(0) $\propto$ $N_S$ $vs.$ $P_{pl}$. \cite{san04,uem89,pum90,zim95} While there is a slight scattering in the magnitude among the reported $\sigma$(0), each reported $\sigma$(0) data set have similar $P_{pl}$-dependence. The $N_S$ linearly increases with doping for $\sim$1/16 ($P_c$) $<$ $P_{pl}$ $<$ $\sim$1/8 (2$P_c$) followed by a clear plateau for $\sim$1/8 (2$P_c$) $<$ $P_{pl}$ $<$ $\sim$3/16 (3$P_c$). It then linearly increases for $\sim$3/16 (3$P_c$) $<$ $P_{pl}$ $<$ $\sim$1/4 (4$P_c$) again, and ends with a sharp peak at $\sim$1/4 (4$P_c$). Finally, the $N_S$ rapidly decreases for $P_{pl}$ $>$ 1/4 (4$P_c$). Therefore, in superfluid density $vs.$ $P_{pl}$ plot, there is also a flat two-phase region bounded by $P_{pl}$ =1/8 and 3/16, consistent with the co-existence of two PEP's of 2$P_c$ and 3$P_c$. It is interesting to point out that the famous linear $T_c$ $vs.$ $N_S$ plot, the Uemura plot, failed to reveal the two phase region. \cite{uem89} Therefore in order to have a physically meaningful comparison of various physical properties of HTS, it is of critical importance that the physical properties should always be plotted in terms of $P_{pl}$ determined by the universal hole-scale. \cite{hon04}

In summary, we have examined both the electronic-phase and chemical-phase diagrams of double-layer high temperature superconductors YBa$_2$Cu$_3$O$_{6+\delta}$ as a function of the hole content per CuO$_2$ plane $P_{pl}$ and oxygen doping concentration $\delta$, respectively. The $T_c$($P_{pl}$), irregardless of the sample quality, is a universal curve originating at $P_c$ = 1/16. The 60K-plateau, $T_c$-jump and 90K-plateau occur at a series hierarchical doping concentration based on $P_c$ such as $P_{pl}$ = 2$P_c$, 3$P_c$ and 4$P_c$. Our analysis suggests that the electronic phase diagram of $T_c$ $vs.$ $P_{pl}$ can be understood, although modified by the chemical phases, in term of the existence and the EPS of PEP's. We conclude that there are at least two PEP's with $T_c$ = 60 K at $P_{pl}$ = 2$P_c$ = 1/8 and $T_c$ = 90 K at $P_{pl}$ = 3$P_c$ = 3/16. Beyond $P_{pl}$ = 4$P_c$ =1/4, the superconductivity, such as $T_c$ and $N_S$, is always suppressed. The observation of superconducting transitions at magic doping levels in the double-layer OD-Y123 and single-layer La-214 strongly suggest that the PEP's are generic intrinsic properties of all high temperature superconductors.

One of us (T.H.) would like to thank Dr. M. Tanimoto of Asahikawa Medical College for offering convincement for study. This work was supported by the state of Texas through the Texas Center for Superconductivity at the University of Houston.

\end{document}